\documentclass{article}
\usepackage{amssymb,epsfig,graphicx}
\usepackage{amsmath, graphics}
\newcommand{\be}{\begin{equation}}
\newcommand{\ee}{\end{equation}}
\newcommand{\bea}{\begin{eqnarray}}
\newcommand{\eea}{\end{eqnarray}}
\newcommand{\barr}{\begin{array}}
\newcommand{\earr}{\end{array}}
\newtheorem{thm}{Theorem}[section]
\newtheorem{defn}{Definition}[section]
\newtheorem{prop}{Proposition}[section]
\newtheorem{rem}{Remark}[section]

\newcommand{\bpar}{\be \left\{ \begin{array}{lll}}
\newcommand{\epar}{ \end{array}\right. \ee }
\newcommand{\eparn}{ \end{array} \right.}

\newcommand\leftmat{\left(\begin{array}{cc}}
\newcommand\rightmat{\end{array}\right)}
\newcommand\leftvec{\left(\begin{array}{c}}
\newcommand\rightvec{\end{array}\right)}
\newcommand\re{{\rm e}}
\newcommand\la{{\lambda}}
\newcommand\R{{\mathbb R}}
\newcommand\C{{\mathbb C}}

\newcommand\RH{Riemann-Hilbert }

\begin{document}
\bibliographystyle{plain}

\vspace{3mm}
\begin{center}
\Large
{\bf Boundary value problems for the elliptic
sine-Gordon equation in a semi-strip}

\vspace{3mm}

\large
A.S. Fokas$^{*}$ and B. Pelloni$^{**}$

\vspace{5mm}

*Department of Applied Mathematics and Theoretical Physics

Cambridge University

Cambridge CB3 0WA, UK.

{\em t.fokas@damtp.cam.ac.uk}

\vspace{2mm}

**Department of Mathematics

University of Reading

Reading RG6 6AX, UK

{\em b.pelloni@reading.ac.uk}

\vspace{3mm}
\today
\end{center}
\normalsize

\begin{abstract}
We study boundary value problems posed in a semistrip for the elliptic sine-Gordon equation, which is the paradigm of an elliptic integrable PDE in two variables. We use the method introduced by one of the authors, which provides a substantial generalization of  the inverse scattering transform and  can be used for the analysis of boundary as opposed to initial-value problems. We first express the solution in terms of a $2\times 2$ matrix \RH problem formulated in terms of both the Dirichlet and the Neumann boundary values on the boundary of a semistrip. We then concentrate on the case that the prescribed boundary conditions are zero along the unbounded sides of the semistrip  and constant along the bounded side; in this particular case we show that the ``jump matrices" of the above \RH problem can be expressed explicitly in terms of the width of the semistrip and the constant value of the solution along the bounded side. This \RH problem has  a unique solution.

\end{abstract}

\section{Introduction}

A method for solving initial-boundary value problems for linear and integrable nonlinear PDEs was introduced in \cite{fok} and developed by several authors \cite{book}. This method has already been used for:

(a) linear and integrable nonlinear {\em evolution} PDEs formulated on the half line and on a finite interval \cite{161,178,179,162,110,cmp,fokIMA,FI1992,177,fis,  fpIMA, FLKdV,p1,p2,TF2008};

(b) linear and integrable nonlinear {\em hyperbolic} PDEs \cite{fokmen,pelpin};

(c) {\em linear elliptic} PDEs \cite{bAF, CF, Dasfok,  best, FK, fokspe,  FZ,111,EF1, EF2}.

The aim of this paper is to implement this method in the case of the prototypical integrable {\em nonlinear elliptic} PDE, namely the celebrated sine-Gordon equation ( simple boundary value problems for this equation, using the method of \cite{fok}, have been analyzed in \cite{PelGal, PP}).

We will consider the sine-Gordon equation in the form
\be
q_{xx}+q_{yy}=\sin q,\quad q=q(x,y),
\label{esg}\ee
and we will analyze boundary value problems posed in the semi-infinite strip
$$
{\cal S}=\{0<x<\infty,\qquad 0<y<L\},
$$
where $L$ is a positive finite constant.
The sides $\{y=L, \;0<x<\infty\}$, $\{x=0, \;0<y<L\}$ and $\{y=0, \;0<x<\infty\}$ will be referred to as side (1), (2) and (3) respectively, see figure 1.

\begin{figure}[t]
  \begin{center}
				      \setlength{\unitlength}{0.04in}
				      \begin{picture}(90,40)(-50,10)
				    \thinlines
				 \put(-50,30){\vector(1,0){95}}
				\put(-50,30){\line(0,1){20}}
				\put(-50,50){\line(1,0){90}}
				\put(-50,50){\vector(0,1){20}}
			\put(-10,40){${\cal S}$}
			\put(-51,72){$y$}
			\put(46,29){$x$}
			\put(-54,49){$L$}
				\put(-55,40){$(2)$}
				\put(-15,53){$(1)$}
				\put(-15,25){$(3)$}
			
\end{picture}
\caption{The semistrip {\cal S}}
\end{center}
\end{figure}
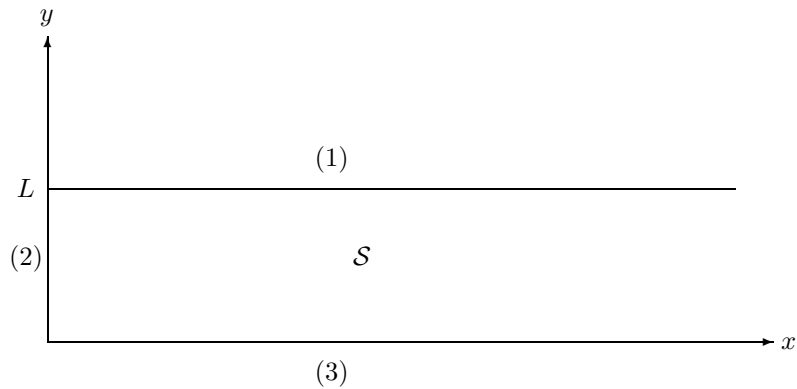

\smallskip
Suppose that (\ref{esg}) is supplemented with appropriate boundary conditions on the boundary of the semistrip ${\cal S}$, so that there exists a unique solution $q(x,y)$. It will be shown in section 2 that this solution can be expressed in terms of the solution of a $2\times 2$ matrix Riemann-Hilbert (RH)  problem with jumps on the union of the real and imaginary axis of the $\lambda$ complex plane.  The "jump matrices"  are expressed in terms of certain functions, called {\em spectral functions}, which will be denoted by $\{a_j(\lambda), b_j(\lambda)\}$, $j=1,2,3$. These functions can be uniquely characterized  via the solution of certain linear Volterra integral equations, in terms of the Dirichlet  and Neumann boundary values. Namely, $\{a_1, b_1\}$, $\{a_2,b_2\}$ and $\{a_3,b_3\}$ are uniquely determined in terms of $\{q(x,L), q_y(x,L)\}$, $\{q(0,y), q_x(0,y)\}$ and $\{q(x,0), q_y(x,0)\}$ respectively.  However, for a well posed problem only a subset of these boundary values are prescribed as boundary conditions. Thus, in order to compute the spectral functions in terms of the given boundary conditions, one must first determine the unknown boundary values, i.e. one must characterize the Dirichlet to Neumann map.
The solution of this problem, which makes crucial use of the so-called {\em global relation}, yields in general a nonlinear map.

In the case of integrable nonlinear evolution PDEs,  it has been shown in \cite{cmp, FoksG,FLgnls,FLKdV} that there exists a particular class of boundary conditions, called linearizable, for which it is possible to avoid the above nonlinear map. The main result of the present paper is the analysis of linearizable boundary conditions for the sine-Gordon equation on the semi-infinite strip. In particular, the following boundary conditions will be investigated in detail:
\be
q(x,L)=q(x,0)=0,\;\;0<x<\infty;\qquad q(0,y)=d, \;\;0<y<L,
\label{lbc}\ee
where $d$ is a finite constant. It will be shown in section 5 that in this particular case, the "jump matrices" of the associated RH problem can be constructed explicitly in terms of the given constant $d$ and of the width $L$ of the semistrip.  This result, as well as the analogous result valid for the elliptic version of the Ernst equation \cite{fernst}, imply that the new method of \cite{fok} provides a powerful tool for analyzing effectively a large class of interesting boundary conditions.

\section{Spectral analysis under the assumption of existence}
\setcounter{equation}{0}
In what follows we assume that (\ref{esg}) is supplemented with appropriate boundary conditions on the boundary of the semistrip ${\cal S}$ so that the existence of a unique solution $q(x,y)$ can be assumed.
Furthermore, we assume the following:
\bea
&&q(x,L), \;q_y(x,L), \;q(x,0), \;q_y(x,0)\in {\rm L}^1(\R^+),
\nonumber \\
&&xq(x,L), \;xq_y(x,L), \;xq(x,0), \;xq_y(x,0)\in {\rm L}^1(\R^+),
\label{bcass}\\
&&q(0,y),\; q_x(0,y), \; yq(0,y),\;yq_x(0,y)\in {\rm L}^1([0,L]).
\nonumber \eea
The sine-Gordon equation is the compatibility condition of the following Lax pair  for the $2\times 2 $ matrix-valued function $\Psi(x,y,\lambda)$, $\lambda\in\C$:
\bea
     &&  \Psi_{x}+\frac{\Omega(\lambda)}{2}[\sigma_{3},\Psi]=Q(x,y,\lambda)
      \Psi,
       \label{lax1mux}\\
&&       \Psi_{y}+\frac{\omega(\lambda)}{2}[\sigma_{3},\Psi]=iQ(x,y,-\lambda)
	  \Psi,
       \label{lax2muy}
       \eea
       where
       \be
       \Omega(\lambda)=\frac 1 {2i}\left(\lambda-\frac 1 \lambda\right),\qquad
        \omega(\lambda)=\frac 1 {2}\left(\lambda+\frac 1 \lambda\right),
        \label{omedef}\ee
       \be
       Q(x,y,\lambda)=     \frac i 4 \left(\begin{array}{cc}
     \frac 1{\lambda}(1-\cos q)&q_x-iq_y+\frac{i\sin q}{\lambda}
     \\ \\
     q_x-iq_y-\frac{i\sin q}{\lambda}&- \frac 1{\lambda}(1-\cos q)
     \end{array}\right),\qquad q=q(x,y).
     \label{Q0def}
     \ee
Equations (\ref{lax1mux}) and  (\ref{lax2muy}) can be written as the single equation
\be
d\left(\re^{(\Omega(\lambda)x+\omega(\lambda)y)\frac{\widehat{\sigma_3}}{2}}\right)\Psi(x,y,\lambda)=W(x,y,\lambda),
\label{dform}\ee
where the differential form $W$ is given by
\be
W(x,y,\lambda)=\re^{(\Omega(\lambda)x+\omega(\lambda)y)\frac{\widehat{\sigma_3}}{2}}
\left(Q(x,y,\lambda)\Psi(x,y,\la)dx+iQ(x,y,-\lambda)\Psi(x,y,\la)dy\right),
\label{Wdef}\ee
and $\widehat{\sigma_3}$ acts on a $2\times 2 $ matrix $A$ by
$$
\widehat{\sigma_3}A=[\sigma_3,A].
$$
\begin{rem}
Note that
$$
\overline{\Omega(\bar\la)}=-\Omega(\la)=\Omega(\frac 1 \la),\qquad
\overline{\omega(\bar\la)}=\omega(\la)=\omega(\frac 1 \la).
$$
\end{rem}
\subsection{Bounded and analytic eigenfunctions}
We define three solutions $\Psi_j(x,y,\lambda)$ $j=1,2,3,$ of (\ref{dform}) by
\be
\Psi_j(x,y,\lambda)=I+\int_{(x_j,y_j)}^{(x,y)}\re^{-(\Omega(\lambda)x+\omega(\lambda)y)\frac{\widehat{\sigma_3}}{2}}W(\xi,\eta,\lambda),
\label{psii}\ee
where
\be
(x_1,y_1)=(\infty,y),\quad (x_2,y_2)=(0,L),\quad (x_3,y_3)=(0,0).
\label{points}\ee

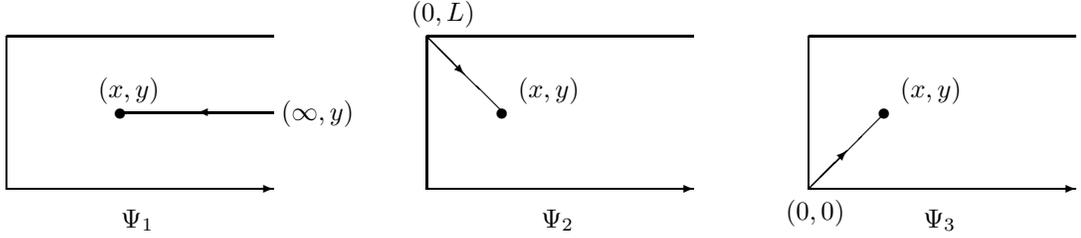
\begin{figure}
\begin{center}
 \setlength{\unitlength}{0.04in}
				      \begin{picture}(90,40)(-50,10)
				    \thinlines
				 \put(-80,20){\vector(1,0){35}}
				\put(-80,20){\line(0,1){20}}
				\put(-80,40){\line(1,0){35}}
				\put(-45,30){\vector(-1,0){10}}
				\put(-55,30){\line(-1,0){10}}
				\put(-68,32){$(x,y)$}
				 \put(-66,29){$\bullet$}
				 \put(-44,29){$(\infty,y)$}
				 \put(-65,15){$\Psi_1$}
				 \put(-25,20){\vector(1,0){35}}
				\put(-25,20){\line(0,1){20}}
				\put(-25,40){\line(1,0){35}}
				\put(-25,40){\line(1,-1){10}}
				\put(-25,40){\vector(1,-1){5}}
				\put(-13,32){$(x,y)$}
				 \put(-16,29){$\bullet$}
				 \put(-27,42){$(0,L)$}
				  \put(-10,15){$\Psi_2$}
 \put(25,20){\vector(1,0){35}}
				\put(25,20){\line(0,1){20}}
				\put(25,40){\line(1,0){35}}
					\put(37,32){$(x,y)$}
					\put(25,20){\line(1,1){10}}
				\put(25,20){\vector(1,1){5}}
				 \put(34,29){$\bullet$}
				 	\put(22,16){$(0,0)$}
				  \put(40,15){$\Psi_3$}

\end{picture}
\caption{The functions $\Psi_i$}
\end{center}
\end{figure}

Since the differential form $W$ is exact, the integral on the right hand side of (\ref{psii}) is independent of the path of integration. We choose the particular contours shown in figure 2. This choice implies the following inequalities on the contours:
$$
(x_1,y_1)\to (x,y):\quad \xi-x\geq 0.
$$
$$
(x_2,y_2)\to (x,y):\quad \xi-x\leq 0,\; \eta-y\geq 0.
$$
$$
(x_3,y_3)\to (x,y):\quad \xi-x\leq 0,\; \eta-y\leq 0.
$$
The first inequality above implies that the exponential appearing in the second (first) column of the right hand side of the equation defining $\Psi_1$ is bounded and analytic for $Im(\lambda)<0$ ($Im(\lambda)>0$). Similar considerations are valid for $\Psi_2$ and $\Psi_3$. Hence we denote the matrices $\Psi_j$ as follows:
$$
\Psi_{1}=(\Psi_{1}^{(12)}, \Psi_{1}^{(34)}),\qquad
\Psi_{2}=(\Psi_{2}^{(4)}, \Psi_{2}^{(2)}),\qquad
\Psi_{3}=(\Psi_{3}^{(3)}, \Psi_{3}^{(1)}),
$$
where the superscript $(12)$ denotes the union of the first and second quadrants of the $\lambda$ complex plane, and similarly for the other superscripts.
The function $\Psi_1^{(12)}$ is analytic for $Im(\lambda)>0$ and it has essential singularities at $\lambda=\infty$ and $\lambda=0$; furthermore,
\be
\Psi_1^{(12)}=\left(\begin{array}{c} 1\\0\end{array}\right)+O\left(\frac 1 \lambda\right),\qquad \lambda\to\infty,\quad Im(\lambda)\geq 0.
\label{as1}\ee
Similar considerations are valid for the column vectors $\Psi_1^{(34)}$, $\Psi_3^{(3)}$ and $\Psi_3^{(1)}$. The function $\Psi_2$ is an analytic function in the entire complex plane, except at $\lambda=\infty$ and $\lambda=0$, where it has essential singularities.  In addition,
\bea
\Psi_2^{(4)}=\left(\begin{array}{c} 1\\0\end{array}\right)+O\left(\frac 1 \lambda\right),&&\quad \lambda\to\infty,\quad \frac {3\pi}{2}<arg(\lambda)<2\pi,
\nonumber\\
\Psi_2^{(2)}=\left(\begin{array}{c} 0\\1\end{array}\right)+O\left(\frac 1 \lambda\right),&&\quad \lambda\to\infty,\quad \frac {\pi}{2}<arg(\lambda)<\pi.
\label{as2}\eea

\subsection{Spectral functions}
Any two solutions $\Psi$, $\tilde{\Psi}$ of (\ref{dform}) are related by an equation of the form
\be
\Psi(x,y,\la)=\tilde\Psi(x,y,\la)\re^{-(\Omega(\lambda)x+\omega(\lambda)y)\frac{\widehat{\sigma_3}}{2}}C(\lambda).
\label{relpsi}\ee
We introduce the notations
\be
S_1(\lambda)=\Psi_1(0,L,\lambda),\quad S_2(\lambda)=\Psi_2(0,0,\lambda),\quad
S_3(\lambda)=\Psi_1(0,0,\lambda).
\label{sf1}\ee
Then equation (\ref{relpsi}) implies the following equations:
\bea
&&\Psi_{1}(x,y,\lambda)=\Psi_{2}(x,y,\lambda)\re^{-(\Omega(\lambda)x+\omega(\lambda)y)\frac{\widehat{\sigma_3}}{2}}\re^{\frac{\omega(\lambda)}{2}L\widehat{\sigma_{3}}}
S_1(\lambda),
\quad
\lambda\in(\R^{+},\R^{-}),
\label{jump1}\\
&&
\Psi_{2}(x,y,\lambda)=\Psi_{3}(x,y,\lambda)\re^{-(\Omega(\lambda)x+\omega(\lambda)y)\frac{\widehat{\sigma_3}}{2}}S_2(\lambda),
\quad
\lambda\in(i\R^{-},i\R^{+}),
\label{jump2}
\\
&&
\Psi_{1}(x,y,\lambda)=\Psi_{3}(x,y,\lambda)\re^{-(\Omega(\lambda)x+\omega(\lambda)y)\frac{\widehat{\sigma_3}}{2}}S_3(\lambda),
\quad
\lambda\in(\R^{-},\R^{+}).
\label{jump3}
\eea
The notation $\lambda\in(\R^{+},\R^{-})$ means that  the equation for first column vector of (\ref{jump1})  is valid for $\lambda\in\R^+$, while the equation for the second vector is valid for $\R^-$, and similarly for (\ref{jump2}), (\ref{jump3}).

Equations (\ref{sf1})-(\ref{jump3}) suggest the following definitions:
\bea
S_1(\lambda)=\Phi_1(0,\lambda),&&
\Phi_1(x,\lambda)=I-\int_x^{\infty}\re^{\Omega(\lambda)(\xi-x)\frac{\widehat{\sigma_3}}{2}}
Q(\xi,L,\lambda)\Phi_1(\xi,\lambda)d\xi,
\nonumber\\
&&\lambda\in(\C^+,\C^-),\quad 0<x<\infty,
\label{psi1}
\\
\nonumber\\
S_2(\lambda)=\Phi_2(0,\lambda),&&
\Phi_2(y,\lambda)=I-i\int_y^{L}\re^{\omega(\lambda)(\eta-y)\frac{\widehat{\sigma_3}}{2}}
Q(0,\eta,-\lambda)\Phi_2(\eta,\lambda)d\eta,
\nonumber\\
&&\lambda\in\C,\quad 0<y<L,
\label{psi2}\\
\nonumber\\
S_3(\lambda)=\Phi_3(0,\lambda),&&
\Phi_3(x,\lambda)=I-\int_x^{\infty}\re^{\Omega(\lambda)(\xi-x)\frac{\widehat{\sigma_3}}{2}}
Q(\xi,0,\lambda)\Phi_3(\xi,\lambda)d\xi,
\nonumber\\
&&\lambda\in(\C^+,\C^-),\quad 0<x<\infty.
\label{psi3}\eea

The matrix $Q$ satisfies the symmetry properties
$$
Q(\lambda)_{22}=Q(-\lambda)_{11},
\qquad Q(\lambda)_{12}=Q(-\lambda)_{21}.
$$
Hence the matrices $\Phi_i$ can be represented in the form
$$
\Phi_1=\left(\begin{array}{cc}
A_{1}(x,\lambda)&B_{1}(x,-\lambda)
\\
B_{1}(x,\lambda)&A_{1}(x,-\lambda)
\end{array}\right),\quad
\Phi_2=\left(\begin{array}{cc}
A_{2}(y,\lambda)&B_{2}(y,-\lambda)
\\
B_{2}(y,\lambda)&A_{2}(y,-\lambda)
\end{array}\right),\quad
\Phi_3=\left(\begin{array}{cc}
A_{3}(x,\lambda)&B_{3}(x,-\lambda)
\\
B_{3}(x,\lambda)&A_{3}(x,-\lambda)
\end{array}\right),
$$
and therefore
$$
S_i(\lambda)=\left(\begin{array}{cc}
a_{i}(\lambda)&b_{i}(-\lambda)
\\
b_{i}(\lambda)&a_{i}(-\lambda)
\end{array}\right),\qquad i=1,2,3.
$$

The spectral functions  $\{a_1(\lambda),\;b_1(\lambda)\}$, $\{a_2(\lambda),\;b_2(\lambda)\}$ and $\{a_3(\lambda),\;b_3(\lambda)\}$ are defined in terms of $\{q(x,L), \;q_y(x,L)\}$, $\{q(0,y), \;q_x(0,y)\}$ and $\{q(x,0), \;q_y(x,0)\}$ respectively, through equations (\ref{psi1})-(\ref{psi3}).

\medskip
These functions have the following properties:
\begin{itemize}
\item

$a_1(\lambda)$, $b_1(\lambda)$ are analytic and bounded in $\C^+$.

$a_1(\lambda)a_1(-\lambda)-b_1(\lambda)b_1(-\lambda)=1$.

$
a_1(\lambda)=1+O\left(\frac 1 \lambda\right)$, $b_1(\lambda)=O\left(\frac 1 \lambda\right)$ as
$\lambda\to\infty,\;Im(\lambda)\geq 0.
$
\item
$a_2(\lambda)$, $b_2(\lambda)$ are analytic functions of $\lambda$ for all $\lambda\in\C$, except for essential singularities at $\lambda=\infty$ and $\lambda=0$.

$a_2(\lambda)a_2(-\lambda)-b_2(\lambda)b_2(-\lambda)=1$.

$
a_2(\lambda)=1+O\left(\frac 1 \lambda\right)$, $b_2(\lambda)=O\left(\frac 1 \lambda\right)$ as
$\lambda\to\infty, \;\frac {3\pi}{2}<arg(\lambda)<2\pi.
$
\item
$a_3(\lambda)$, $b_3(\lambda)$ are analytic and bounded in $\C^+$.

$a_3(\lambda)a_3(-\lambda)-b_3(\lambda)b_3(-\lambda)=1$.

$
a_3(\lambda)=1+O\left(\frac 1 \lambda\right)$, $b_3(\lambda)=O\left(\frac 1 \lambda\right)$ as
$\lambda\to\infty,\;Im(\lambda)\geq 0.
$
\end{itemize}

These properties follow from the analogous properties of the matrix-valued functions $\Phi_j$, $j=1,2,3$, from the condition of unit determinant, and from the large $\lambda$ asymptotics of these functions.

\subsection{The global relation}
Evaluating equations (\ref{jump2}) and (\ref{jump3}) $at\;x=0,\,y=L$, we find
$$
I=\Psi_3(0,L,\lambda)\re^{-\frac{\omega(\lambda)}{2}L\widehat{\sigma_{3}}}
S_2(\lambda)
$$ and
$$
S_{1}(\lambda)=\Psi_{3}(0,L,\lambda)\re^{-\frac{\omega(\lambda)}{2}L\widehat{\sigma_{3}}}
S_3(\lambda).
$$
Eliminating $\Psi_{3}(0,L,\lambda)$ we obtain
\be
\re^{\frac{\omega(\lambda)}{2}L\widehat{\sigma_{3}}}S_{1}(\lambda)=S_2(\lambda)^{-1}S_3(\lambda).
\label{gr0.0}\ee
The first column vector of this equation yields the following {\em global relations}:
\bea
 &a_{1}(\lambda)=a_{2}(-\lambda) a_{3}(\lambda)-b_2(-\lambda)b_3(\lambda),
& \qquad \lambda\in \C^+,
 \label{grab1a}\\
 &b_{1}(\lambda)\re^{-\omega(\lambda)L}=a_{2}(\lambda) b_{3}(\lambda)-a_3(\lambda)b_2(\lambda),&\qquad \lambda\in \C^+.
\label{grab2a}\eea

\subsection{The Riemann-Hilbert problem}\label{basicrh}
Equations (\ref{jump1})-(\ref{jump3}),relating the various analytic
eigenfunctions, can be rewritten in a form that determines the jump conditions of  a $2\times 2$
RH problem, with unitary jump matrices on the real and imaginary axes. This involves tedious but straightforward algebraic manipulations. The final form is
\be
\Psi_{-}(x,y,\lambda)=\Psi_{+}(x,y,\lambda)J(x,y,\lambda), \lambda\in\R\cup i\R,
\label{rh}\ee
where  the matrices $\Psi_{\pm}$ and $J$ are defined as follows:
\bea
\Psi_{+}&=&\left(\Psi_{1}^{(12)},
\frac {1}{a_3(\lambda)}\Psi_{3}^{(1)}\right),\quad
arg(\lambda)\in[0,\frac{\pi}{2}],
\nonumber \\
\Psi_{-}&=&\left(\begin{array}{c}
\Psi_{1}^{(12)},\frac {1}{a_1(\lambda)}\Psi_{2}^{(2)}\end{array}\right),\quad
arg(\lambda)\in[\frac{\pi}{2},\pi],
\nonumber \\
\Psi_{+}&=&\left(\frac {1}{a_3(-\lambda)}\Psi_{3}^{(3)},\Psi_{1}^{(34)}\right),\quad
arg(\lambda)\in[\pi,\frac{3\pi}{2}],
\nonumber \\
\Psi_{-}&=&\left(\begin{array}{c}
\frac {1}{a_1(-\lambda)}\Psi_{2}^{(4)},\Psi_{1}^{(34)}\end{array}\right),\quad
arg(\lambda)\in[\frac{3\pi}{2},2\pi],
\nonumber
\eea

\be
J(x,y,\lambda)=
J^{\alpha}(x,y,\lambda),\;\;{\rm if}\;\;arg(\lambda)=\alpha,\quad
\alpha=0,\,\frac{\pi}{2},\, \pi,\,\frac{3\pi}{2}, \ee
where, using the global relation, we find
$$
J^{0}=\left(\begin{array}{cc}
\frac {a_2(\lambda)}{a_1(-\lambda)a_3(\lambda)}& \frac {b_3(-\lambda)}{a_3(\lambda)}\re^{-\theta(x,y,\lambda)}
\\
\\
-\frac{\re^{-\omega(\lambda)L}b_1(\lambda)}{a_1(-\lambda)}\re^{\theta(x,y,\lambda)}& 1
 \end{array}\right),
$$

 $$
 J^{\pi/2}=\left(\begin{array}{cc}
1&
\frac{b_2(-\lambda)}{a_1(\la)a_3(\lambda)}\re^{-\theta(x,y,\lambda)}
\\
\\
0&
1
 \end{array}\right),\qquad
 J^{3\pi/2}=\left(\begin{array}{cc}
1& 0
\\
\\
\frac{b_2(\lambda)}{a_1(-\lambda)a_3(-\lambda)}\re^{\theta(x,y,\lambda)}&1
 \end{array}\right)
$$
and
\be
J^{\pi}=
J^{3\pi/2}(J^{0})^{-1}J^{\pi/2},
\label{alljumps}\ee
where
\be
\theta(x,y,\lambda)=\Omega(\lambda)x+\omega(\lambda)y.
\label{Theta}\ee
All the matrices $J^{\alpha}$ have unit
determinant: for $J^{\pi/2}$ and $J^{3\pi/2}$ this is immediate, whereas for $J^0$ we find
$$
\det(J^0)=\frac{a_2(\lambda)+\re^{-\omega(\lambda)L}b_1(\lambda)b_3(-\lambda)}{a_1(-\lambda)a_3(\lambda)}=
\frac {a_1(-\lambda)a_3(\lambda)}{a_1(-\lambda)a_3(\lambda)}=1,
$$
where we have used the equation
\be
a_2(\lambda)=a_1(-\la)a_3(\la)-b_3(-\la)b_1(\la)\re^{-\omega(\lambda)L},\qquad \la\in\R.
\label{gfordet}\ee
Equation (\ref{gfordet}) is a consequence of equations (\ref{grab1a}) and (\ref{grab2a}) (see also equation (\ref{rel3}) below).


\medskip
\begin{figure}[h]
				  \begin{center}
				      \setlength{\unitlength}{0.04in}
				      \begin{picture}(90,40)(-50,10)
				    \thinlines
				 \put(-10,30){\line(1,0){30}}
				 \put(-10,30){\line(-1,0){30}}
				\put(-10,30){\line(0,1){20}}
				\put(-10,30){\line(0,-1){20}}
\put(-10,40){\vector(0,-1){2}}
\put(0,30){\vector(1,0){2}}
\put(-10,20){\vector(0,1){2}}
\put(-20,30){\vector(-1,0){2}}
	\put(-4,19){$(\frac 1 {a_1(-\lambda)}\Psi_2^{(4)}, \Psi_{1}^{(34))})$}
	\put(-4,39){$(\Psi_{1}^{(12)},
\frac {1}{a_3(\lambda)}\Psi_{3}^{(1)})$}
	\put(-39,39){$(\Psi_{1}^{(12)}, \frac 1{a_1(\lambda)}\Psi_{2}^{(2)})$}
	\put(-39,19){$(\frac {1}{a_3(-\lambda)}\Psi_{3}^{(3)},\Psi_{1}^{(34)})$}
\put(22,30){$J^{0}$}
	\put(-45,30){$J^{\pi}$}
	\put(-14,52){$J^{\pi/2}$}
	\put(-14,6){$J^{3\pi/2}$}
	\put(-9,31){$\Gamma$}
	
				   \end{picture}
				   \caption{Bounded eigenfunctions
				   and the Riemann-Hilbert problem}
			       \end{center}
				
    \label{fig1}
    \end{figure}
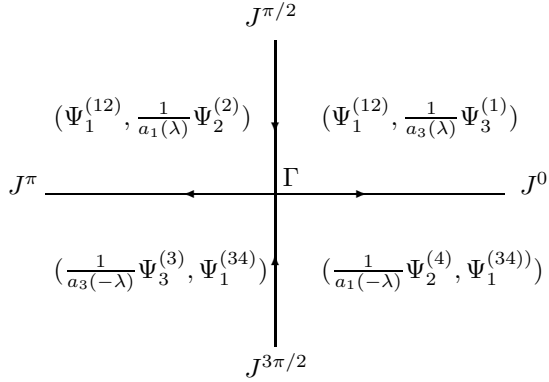

The function $\Psi(x,y,\la)$ solution of this RH problem is a sectionally meromorphic function of $\la$. The possible poles of this function are generated by the zeros of the function $a_1(\la)$ in the region $\{arg(\la)\in[\frac{\pi}{2},\pi]\}$, by the zeros of $a_3(\la)$ in the region $\{arg(\la)\in[0,\frac\pi 2]\}$,  and by the corresponding zeros of $a_1(-\la)$, $a_3(-\la)$.

\smallskip
We assume
\begin{itemize}
\item
The possible zeros of $a_1$  in the region $\{arg(\la)\in[\frac{\pi}{2},\pi]\}$  are simple; these zeros are denoted $\la_j$, $j=1,..,N_1$
\be\label{zeros}\ee
\item
The possible zeros of $a_3$  in the region $\{arg(\la)\in[0,\frac\pi 2]\}$  are simple; these zeros are denoted $\zeta_j$, $j=1,..,N_3$
\end{itemize}

The residues of the function $\Psi$ at the corresponding poles can be computed using equations (\ref{jump1})-(\ref{jump3}). Indeed, equation (\ref{jump3}) yields
$$
\Psi_1^{(12)}=a_3\Psi_3^{(3)}+b_3\re^{\theta(x,y,\la)}\Psi_3^{(1)},
$$
hence
\be
Res_{\zeta_j} \frac{\Psi_3^{(1)}}{a_3}=\frac{\Psi_3^{(1)}(\zeta_j)}{\dot a_3(\zeta_j)}=\frac{\Psi_1^{(12)}(\zeta_j)}{\dot a_3(\zeta_j)b_3(\zeta_j)}\re^{-\theta(x,y,\zeta_j)},
\label{resa3}\ee
where
$
\dot a_3(\la)$ denotes the derivative of $a_3$ with respect to $\la$.

Similarly, using (\ref{jump1}),
\be
Res_{\la_j} \frac{\Psi_2^{(2)}}{a_1}=\frac{\Psi_2^{(2)}(\la_j)}{\dot a_1(\la_j)}=\frac{\Psi_1^{(12)}(\la_j)}{\dot a_1(\la_j)b_1(\la_j)\re^{-\omega(\la_j)L}}\re^{-\theta(x,y,\la_j)}.
\label{resa1}\ee

Using the notation $[\Psi]_1$ for the first column, $[\Psi]_2$ for the second column  for the solution $\Psi$ of the RH problem (\ref{rh}),
at equations (\ref{resa3}) and (\ref{resa1}) imply the following residue conditions:
\bea
Res_{\zeta_j}[\Psi(x,y,\la)]_2&=&\frac{\re^{-\theta(x,y,\zeta_j)}}{\dot a_3(\zeta_j)b_3(\zeta_j)}[\Psi(x,y,\zeta_j)]_1,
\qquad 0<\arg\la<\frac{\pi}{2},
\nonumber \\
\label{rescond}
\\
Res_{\la_j}[\Psi(x,y,\la)]_2&=&\frac{\re^{-\theta(x,y,\zeta_j)}}{\dot a_1(\la_j)b_1(\la_j)\re^{-\omega(\la_j)L}}[\Psi(x,y,\la_j)]_1,
\qquad \frac{\pi}{2}<\arg\la<\pi,
\nonumber
\eea
and similar ones in $\C^-$ by letting $\la\to-\la$.


\subsubsection{The inverse problem}
Rewriting the jump condition, we obtain
\be
\Psi_{+}-\Psi_{-}=\Psi_{+}-\Psi_{+}J-=\Psi_{+}(I-J)\Rightarrow
\Psi_{+}-\Psi_{-}=\Psi_{+}\tilde J,
\label{finjump}\ee
where $\tilde J=I-J.$
The asymptotic
 conditions (\ref{as1})-(\ref{as2})) imply
  \be
 \Psi(x,y,\la)=I+\frac{\Psi^{*}(x,y)}{\lambda}+
 O\left(\frac{1}{\lambda^{2}}\right),\quad |\lambda|\to\infty.
\label{longas}
 \ee
 Equations (\ref{finjump}) and (\ref{longas}) define a Riemann-Hilbert problem.

 The solution of this RH problem is given by
 \be
 \Psi(x,y,\lambda)=I+\frac{1}{2\pi i}\int_{\Gamma}\frac{\Psi_{+}(x,y,\lambda')\tilde
 J(x,y,\lambda')}{\lambda'-\lambda}d\lambda',\quad \lambda
\in\Gamma,
\label{repr1}\ee
 where
 $$
 \Gamma=\R\cup i\R.
 $$
 Equations (\ref{longas}) and (\ref{repr1}) imply
 \be
 \Psi^{*}=-\frac{1}{2\pi i}\int_{\Gamma}\Psi_{+}(x,y,\lambda)\tilde
 J(x,y,\lambda)d\lambda.
\label{psiup1}
 \ee

Using (\ref{longas}) in the first ODE in the Lax pair (\ref{lax1mux}), we
 find
 \be
 -\frac{i}{4}[\sigma_{3},\Psi^{*}]=i\frac{q_{x}-iq_y}{4}\sigma_{1}\Rightarrow
 q_{x}-iq_y=2(\Psi^{*})_{21}=2\lim_{\lambda\to\infty}(\lambda \Psi_{21}),
 \label{qexp1}\ee
 ($\sigma_1$, $\sigma_3$ denote the usual Pauli matrices).

In order to obtain an expression in terms of $q$ rather than its derivatives, we
consider the coefficient of the term $\lambda^{-1}$. The (1,1) element of this coefficient yields
 \be    \cos q(x,y)=1+4i(\Psi^*_x)_{11}-2(\Psi^*)^2_{21}.
		       \label{qexp2}\ee

\section{Spectral theory assuming  the validity of the global relation}
\setcounter{equation}{0}

\subsection{The spectral functions}

The above analysis motivates the following definitions for the spectral functions.

\subsubsection{The spectral functions at the $y=0$ and $y=L$ boundaries}
\begin{defn}\label{1def}
Given the functions $q(x,L)$, $q_y(x,L)$ satisfying conditions (\ref{bcass}), define the map
$$
\mathbb S_1:\{q(x,L),q_y(x,L)\}\to \{a_1(\lambda),b_1(\lambda)\}
$$
by
$$
\left(\begin{array}{cc}
a_1(\lambda)\\b_1(\lambda)\end{array}\right)=[\Phi_1(0,L)]_1,\quad \lambda\in\C^+,
$$
where $[\Phi_1(x,L)]_1$ denotes the first column vector of  the unique solution $\Phi_1(x,L)$ of the Volterra linear integral equation
\be
\Phi(x,\lambda)=I-\int_x^{\infty}\re^{\Omega(\lambda)(\xi-x)\frac{\widehat{\sigma_3}}{2}}
Q(\xi,L,\lambda)\Phi(\xi,\lambda)d\xi,
\label{ode1}\ee
and $Q(x,L,\lambda)$ is given in terms of $q(x,L)$ and $q_y(x,L)$ by equation (\ref{Q0def}).
\end{defn}

\begin{prop}\label{1prop}
The spectral functions $a_1(\lambda)$, $b_1(\lambda)$ have the following properties.
\begin{itemize}
\item[(i)]
$a_1(\lambda)$, $b_1(\lambda)$ are continuous and bounded for $Im(\lambda)\geq 0$, and analytic
for $Im(\lambda>0$.
\item[(ii)]
$
a_1(\lambda)=1+O\left(\frac 1 \lambda\right)$, $b_1(\lambda)=O\left(\frac 1 \lambda\right)$ as
$\lambda\to\infty,\;Im(\lambda)\geq 0.
$
\item[(iii)]
$
a_1(\lambda)=\cos\frac {q(x,L)}{2}+O\left( \lambda\right)$, $b_1(\lambda)=i\sin\frac {q(x,L)}{2}+O\left (\lambda\right)$ as
$\lambda\to 0,\;Im(\lambda)\geq 0.
$
\item[(iv)]
$a_1(\lambda)a_1(-\lambda)-b_1(\lambda)b_1(-\lambda)=1$, $Im(\lambda)\geq 0$.
\item[(v)]
The map $\mathbb Q_1:\{a_1,\,b_1\}\to\{q(x,L)\,q_y(x,L)\}$, inverse to $\mathbb S_1$, is given by
$$
 \cos q(x,L)=1+4i\lim_{\la\to\infty}(\la M_x)_{11}+2\lim_{\la\to\infty}(\la M)_{21},
$$
$$
q_y(x,L)=-iq_x(x,L)+2\lim_{\lambda\to\infty}(\lambda M)_{21},
$$
where $M$ is the solution of the  following Riemann-Hilbert problem:
\begin{itemize}
\item[*] The function
$$
M(x,\la)=\left\{\begin{array}{ll}
M_+(x,\la)&\la\in\C^+
\\
M_-(x,\la)&\la\in\C^-
\end{array}\right.
$$
is a sectionally meromorphic function of
 $\la\in\C.$
\item[*]
$M=I+O\left(\frac 1 \la\right)$ as $\la\to\infty$, and
$$
M_+(x,\la)=M_-(x,\la)J_1(x,\la),\qquad \la\in\R,
$$
where
\be
J_1(x,\la)=\left(\begin{array}{cc}
1&-\frac{b_1(-\la)}{a_1(\la)}\re^{-\Omega(\la)x}
\\
\frac{b_1(\la)}{a_1(-\la)}\re^{\Omega(\la)x}&\frac 1 {a_1(\la)a_1(-\la)}
\end{array}\right),\qquad \la\in\R.
\label{jm1}\ee
\item[*]
The function $a_1(\la)$ may have $N_1$ simple poles $\la_j$ in $\C^+$.
\item[*]
Let $[M]_i$ denote the $i$-th column vector of $M$, $1=1,2$. The possible poles of $M_+$ occur at $\la_j$, and the possible poles of $M_-$ occur at $-\la_j$ in $\C^-$, and the associated residues are given by
\bea
Res_{\la_j}[M(x,\la)]_2&=&\frac{\re^{-\Omega(\la_j)x}}{\dot a_1(\la_j)b_1(\la_j)}[M(x,\la_j)]_1,
\nonumber \\
Res_{-\la_j}[M(x,\la)]_1&=&\frac{\re^{\Omega(\la_j)x}}{\dot a_1(-\la_j)b_1(-\la_j)}[M(x,-\la_j)]_2.
\label{res1prob}\eea
\end{itemize}
\end{itemize}
\end{prop}

The spectral functions $\{a_3,b_3\}$ satisfy an analogous result:
\begin{defn}\label{3def}
Given the functions $q(x,0)$, $q_y(x,0)$, satisfying conditions (\ref{bcass}), define the map
$$
\mathbb S_3:\{q(x,0),q_y(x,0)\}\to \{a_3(\lambda),b_3(\lambda)\}
$$
by
$$
\left(\begin{array}{cc}
a_3(\lambda)\\b_3(\lambda)\end{array}\right)=[\Phi_3(0,0)]_1,\quad \lambda\in\C^+,
$$
where $[\Phi_3(x,0)]_1$ denotes the first column vector of  the unique solution $\Phi_3(x,0)$ of the Volterra linear integral equation
\be
\Phi(x,\lambda)=I-\int_x^{\infty}\re^{\Omega(\lambda)(\xi-x)\frac{\widehat{\sigma_3}}{2}}
Q(\xi,0,\lambda)\Phi(\xi,\lambda)d\xi,
\label{ode1.1}\ee
and $Q(x,0,\lambda)$ is given in terms of $q(x,0)$ and $q_y(x,0)$ by equation (\ref{Q0def}).
\end{defn}

\begin{prop}\label{3prop}
The spectral functions $a_3(\lambda)$, $b_3(\lambda)$ have the properties (i)-(v) of proposition (\ref{1prop}), provided $a_1$ is replaced by $a_3$, $b_1$ is replaced by $b_3$,  $
\mathbb S_1$ is replaced by $
\mathbb S_3$ and   $L$ in replaced by  $0$ in all expressions.
\end{prop}

\subsubsection{The spectral functions at the $x=0$ boundary}
\begin{defn}\label{2def}
Given the functions $q(0,y)$, $q_x(0,y)$, satisfying conditions (\ref{bcass}), define the map
$$
\mathbb S_2:\{q(0,y),q_x(0,y)\}\to \{a_2(\lambda),b_2(\lambda)\}
$$
by
$$
\left(\begin{array}{cc}
a_2(\lambda)\\b_2(\lambda)\end{array}\right)=[\Phi_2(0,0)]_1,\quad \lambda\in\C^+,
$$
where $[\Phi_2(0,y)]_1$ denotes the first column vector of  the unique solution $\Phi_2(0,y)$ of the Volterra linear integral equation
\be
\Phi(y,\lambda)=I-i\int_y^L\re^{\omega(\lambda)(\eta-y)\frac{\widehat{\sigma_3}}{2}}
Q(0,y,-\lambda)\Phi(\eta,\lambda)d\eta,
\label{ode1.2}\ee
and $Q(0,y,\lambda)$ is given in terms of $q(0,y)$ and $q_x(0,y)$ by equation (\ref{Q0def}).
\end{defn}

\begin{prop}\label{2prop}
The spectral functions $a_2(\lambda)$, $b_2(\lambda)$ have the following properties.
\begin{itemize}
\item[(i)]
$a_2(\lambda)$, $b_2(\lambda)$ are entire functions of $\la$, except for essential singularities at $\la=0$ and $\la=\infty$,  bounded for $Re(\lambda)> 0$.
\item[(ii)]
$
a_2(\lambda)=1+O\left(\frac 1 \lambda\right)$, $b_2(\lambda)=O\left(\frac 1 \lambda\right)$ as
$\lambda\to\infty,\;Re(\lambda)\geq 0.
$
\item[(iii)]
$
a_2(\lambda)=\cos\frac {q(0,y)}{2}+O\left( \lambda\right)$, $b_2(\lambda)=i\sin\frac {q(0,y)}{2}+O\left (\lambda\right)$ as
$\lambda\to 0,\;Im(\lambda)\geq 0.
$
\item[(iv)]
$a_2(\lambda)a_2(-\lambda)-b_2(\lambda)b_2(-\lambda)=1$, $Im(\lambda)\geq 0$.
\item[(v)]
The map $\mathbb Q_2:\{a_2,\,b_2\}\to\{q(0,y)\,q_y(0,y)\}$, inverse to $\mathbb S_2$, is given by
$$
 \cos q(0,y)=1+4i\lim_{\la\to\infty}(\la M_y)_{11}+2\lim_{\la\to\infty}(\la M)_{21},
$$
$$
q_x(0,y)=iq_y(0,y)+2\lim_{\lambda\to\infty}(\lambda M)_{21},
$$
where $M$ is the solution of the  following Riemann-Hilbert problem:
\begin{itemize}
\item[*] The function
$$
M(y,\la)=\left\{\begin{array}{ll}
M_+(y,\la)&Re\la\geq 0
\\
M_-(y,\la)&Re\la\leq 0
\end{array}\right.
$$
is a sectionally meromorphic function of $\la\in\C.$
\item[*]
$M=I+O\left(\frac 1 \la\right)$ as $\la\to\infty$, and
$$
M_+(y,\la)=M_-(y,\la)J_2(y,\la),\qquad \la\in i\R,
$$
where
$$J_2(y,\la)=\left(\begin{array}{cc}
1&-\frac{b_2(-\la)}{a_2(\la)}\re^{-\omega(\la)x}
\\
\frac{b_2(\la)}{a_2(-\la)}\re^{\omega(\la)x}&\frac 1 {a_2(\la)a_2(-\la)}
\end{array}\right),\qquad \la\in i\R.
$$
\item[*]
$M$ satisfies appropriate residue conditions at the zeros of $a_2(\la)$.
\end{itemize}
\end{itemize}
\end{prop}

\subsubsection{Proof of propositions (\ref{1prop})-(\ref{2prop})}
The proof  of  properties (i)-(iv) follows from the discussion in Section 2.2. In particular, property (iii) follows from the asymptotic behaviour at $\la\to 0$, which can be derived by analysing equations (\ref{lax1mux})-(\ref{lax2muy}) (see \cite{PelGal}), and is given by
\be
\Psi=\Psi_{0}+O\left(\lambda\right),\;
|\lambda|\to 0,\qquad \Psi_0(x,y)=
\left(\begin{array}{cc}
\cos\frac {q(x,y)}{2}&i\sin\frac{q(x,y)}{2}
\\
i\sin\frac{q(x,y)}{2}&\cos\frac {q(x,y)}{2}
\end{array}\right).
\label{asymp0}
\ee

To prove (v), we note that the function $\phi_1(x,\la)$ is the unique solution of the ODE
\bea
&&\phi_x+\frac{\Omega(\la)}{2}\widehat{\sigma_3}\phi=Q(x,L,\la)\phi(x,\la),\nonumber \\
&&\lim_{\la \to\infty}\phi(x,\la)=I.
\nonumber \eea
Furthermore, $\phi_3(x,\la)$ is the solution of the same ODE problem, with $Q(x,L,\la)$ replaced by $Q(x,0,\la)$.

Similarly, $\phi_2(x,\la)$ is the unique solution of the ODE
\bea
&&\phi_y+\frac{\omega(\la)}{2}\widehat{\sigma_3}\phi=iQ(0,y,-\la)\phi(x,\la),\nonumber \\
&&\phi(0,L)=I.
\nonumber \eea

The spectral analysis of the above ODEs yields the desired result.

Regarding the rigorous derivation of the above results, we note the following: If $\{q(x,L), q_y(x,L)\}$, $\{q(x,0), q_y(x,0)\}$  and $\{q(y,0), q_x(y,0)\}$ are in ${\bf L}^1$, then the Volterra integral equations (\ref{ode1}),  (\ref{ode1.1}) and  (\ref{ode1.2}) respectively, have a unique solution, and hence the spectral functions $\{a_j,b_j\}$, $j=1,..,3$, are well defined. Moreover, under the assumption (\ref{bcass}) the spectral functions belong to ${\bf H}^1(\R)$ , hence the \RH problems that determine the inverse maps can be characterized through the solutions of a Fredholm integral equation, see \cite{deift,zhou}.

{\bf QED}

\subsection{The \RH problem}

\begin{thm}
Suppose that a subset of the boundary values $\{q(x,L), q_y(x,L)\},\{q(x,0), q_y(x,0)\}$, $0<x<\infty$, and $\{q(y,0), q_x(y,0)\}$, $0<y<L$, satisfying (\ref{bcass}), are prescribed as boundary conditions. Suppose that these prescribed boundary conditions are such that the global relations (\ref{grab1a}) and (\ref{grab2a}) can be used to characterize the remaining boundary values.

Define the spectral functions $\{a_j,b_j\}$, $j=1,..,3,$ by definitions (\ref{1def})-(\ref{2def}). Assume that the possible zeros $\{\la_j\}_{j=1}^{N_1}$ of $a_1(\la)$ and $\{\zeta_j\}_{j=1}^{N_2}$ of
$a_3(\la)$ are as in assumption \ref{zeros}.

Define $M(x,y,\la)$ as the solution of the following $2\times 2$ matrix \RH problem:

\begin{itemize}
\item[*] The function
$
M(x,y,\la)
$
is a sectionally meromorphic function of
 $\la$ away from $\R\cup i\R$.
 \item[*]
The possible poles of the second column of $M$ occur at $\la=\zeta_j$, $j=1,...,N_2$, in the first quadrant and at $\la=\la_j$, $j=1,...,N_1$, in the second quadrant of the complex $\la$ plane.

The possible poles of the first column of $M$ occur at $\la=-\la_j$ ($j=1,...,N_1$) and $\la=-\zeta_j$ ($j=1,...,N_2$).

The associated residue conditions satisfy the relations (\ref{rescond}).
\item[*]
$M=I+O\left(\frac 1 \la\right)$ as $\la\to\infty$, and
$$
M_+(x,y,\la)=M_-(x,y,\la)J(x,y,\la),\qquad \la\in\R\cup i\R,
$$
where $M=M_+$ for $\la$ in the first or third quadrant, and $M=M_-$ for $\la$  in the second or fourth quadrant of the complex $\la$ plane, and $J$ is defined in terms of $\{a_j,b_j\}$ by equations (\ref{alljumps}).
\end{itemize}

Then $M$ exists and is unique, provided that the ${\bf H}^1$ norm of of the spectral functions is sufficiently small.

\smallskip
Define $q(x,y)$ is terms of $M(x,y,\la)$ by
\bea
q_{x}-iq_y&=&2\lim_{\lambda\to\infty}(\lambda M)_{21},
 \label{inv1}\\
  \cos q(x,y)&=&1+4i(\lim_{\lambda\to\infty}(\lambda M_x)_{11})-2(\lim_{\lambda\to\infty}(\lambda M)_{21})^2.
                                       \label{inv2}\eea
Then $q(x,y)$ solves (\ref{esg}). Furthermore, $q(x,y)$ evaluated at the boundary, yields the functions used for the computation of the spectral functions.
\end{thm}

{\bf Proof}:  Under the assumptions (\ref{bcass}), the spectral functions are in ${\bf H}^1$.

In the case when $a_1(\la)$ and $a_3(\la)$ have no zeros,  the \RH problem is regular and it is equivalent to a Fredholm integral equation. However, we have {\em not} been able to establish a vanishing lemma, hence we require a small norm assumption for solvability.

If $a_1(\la)$ and $a_3(\la)$ have zeros, the singular RH problem can be mapped to  a regular one coupled with a system of algebraic equations \cite{FI96}. Moreover, it follows from standard arguments, using the dressing method \cite{zs1,zs2},
that if $M$ solves the above RH problem and $q(x,y)$ is defined by  (\ref{inv1})-(\ref{inv2}), then $q(x,y)$ solves equation (\ref{esg}). The proof that $q$ evaluated at the boundary yields the functions used for the computation of the spectral functions follows arguments similar to the ones used in \cite{fis}.

{\bf QED}

\section{Linearizable boundary conditions}
\setcounter{equation}{0}
We now concentrate on the particular boundary conditions (\ref{lbc}).

In this case, equations (\ref{psi1})-(\ref{psi3}) simplify as follows:
\bea
\left(\begin{array}{c}
A_1(x,\lambda)\\B_1(x,\lambda)
\end{array}\right)&=&\left(\begin{array}{c}
1\\0\end{array}\right)-\frac 1 4\int_x^{\infty}\left(\begin{array}{c}
q_y(\xi,L)A_1(\xi,\lambda)\\\re^{\Omega(\lambda)(x-\xi)}q_y(\xi,L)B_1(\xi,\lambda)
\end{array}\right)d\xi,
\nonumber \\
&&0<x<\infty,\qquad Im(\lambda)\geq 0,
\label{A1eq}\\
\left(\begin{array}{c}
A_2(y,\lambda)\\B_2(y,\lambda)
\end{array}\right)&=&\left(\begin{array}{c}
1\\0\end{array}\right)+\frac 1 4\int_y^{L}\left(\begin{array}{c}
-\frac{(1-\cos d)}{\lambda}A_2(\eta,\lambda)+[q_x(0,y)-i\frac{\sin d}{\lambda}]B_2(\eta,\lambda)
\\
\re^{\omega(\lambda)(y-\eta)}\left[q_x(0,y)+i\frac{\sin d}{\lambda}]A_2(\eta,\lambda)+
\frac{(1-\cos d)}{\lambda}B_2(\eta,\lambda)\right]
\end{array}\right)d\eta,
\nonumber \\
&&0<y<L,\qquad \lambda\in\C,
\label{A2eq}\\
\left(\begin{array}{c}
A_3(x,\lambda)\\B_3(x,\lambda)
\end{array}\right)&=&\left(\begin{array}{c}
1\\0\end{array}\right)-\frac 1 4\int_x^{\infty}\left(\begin{array}{c}
q_y(\xi,0)A_3(\xi,\lambda)\\\re^{\Omega(\lambda)(x-\xi)}q_y(\xi,0)B_3(\xi,\lambda)
\end{array}\right)d\xi,
\nonumber \\
&&0<x<\infty,\qquad Im(\lambda)\geq 0.
\label{A3eq}
\eea
In equations (\ref{A1eq}) and (\ref{A3eq}), the only dependence on $\lambda$ is through $\Omega(\lambda)$.  Thus, since $\Omega(-\frac 1 \lambda)=\Omega(\lambda)$, it follows that the vector functions $(A_1,B_1)$ and $(A_3,B_3)$ satisfy the same symmetry properties. Hence,
\be
a_j(-\frac 1 \la)=a_j(\la),\quad b_j(-\frac 1 \la)=b_j(\la),\qquad j=1,3, \quad Im(\lambda)\geq 0.
\label{symm1}\ee
It turns out that the vector function $(A_2,B_2)$ also satisfies a certain symmetry condition, as stated in the following proposition.
\begin{prop}
Let $q_x(0,y)$ be a sufficiently smooth function. Then the vector solution of the linear Volterra integral equation (\ref{A2eq}) satisfies the following symmetry conditions (where we do not indicate the explicit dependence of $A_2$, $B_2$ on $y$):
\bea
A_2(\frac 1 \la)=\frac 1 {1-F(\la)^2}[A_2(\la)-F(\la)B_2(\la)+F(\la)\re^{\omega(\la)(y-L)}B_2(-\la)-
F(\la)^2\re^{\omega(\la)(y-L)}A_2(-\la)],
\nonumber \\
B_2(\frac 1 \la)=\frac 1 {1-F(\la)^2}[B_2(\la)-F(\la)A_2(\la)+F(\la)\re^{\omega(\la)(y-L)}A_2(-\la)-
F(\la)^2\re^{\omega(\la)(y-L)}B_2(-\la)],
\nonumber \eea
\be
\\
0<y<L,\quad \la\in\C,\label{symm2}
\ee
where the function $F(\la)$ is given by
\be
F(\la)=i\frac{1-\la^2}{1+\la^2}\tan\frac d 2.
\label{Fdef}\ee

\end{prop}

{\bf Proof}: Let the $2\times 2$ matrix valued function $\Phi_2(y,\lambda)$ be defined by
\be
\Phi_2(y,\la)=\left(\begin{array}{cc}
A_2(y,\la)&B_2(y,-\la)
\\
B_2(y,\la)&A_2(y,-\la)
\end{array}\right),\qquad 0<y<L,\quad \la\in\C.
\label{Phi2}\ee
Then $\Phi_2$ satisfies the ODE
\bea
&&(\Phi_2)_y+\frac {\omega(\la)}{2} [\sigma_3,\Phi_2]=iQ(0,y,-\la)\Phi_2,\qquad 0<y<L,
\nonumber \\
&&\Phi_2(L,\la)=I,
\label{ic}
\eea
where $Q(x,y,\la)$ is defined in (\ref{Q0def}), and $q(0,y)=d$.

Letting
\be
\Phi_2(y,\la)=\phi_2(y,\la)\re^{\frac{\omega(\la)}{2}\sigma_3(y-L)}
\label{Fi2}\ee
it follows that $\phi_2$ satisfies the ODE
\bea
&&(\phi_2)_y=V\phi_2,\label{phi2V} \\
&&\phi_2(L,\la)=I,\qquad 0<y<L,\quad Re(\la)\geq 0,
\nonumber
\eea
where
\be
V(y,\la)=\frac 1 4\left(\begin{array}{cc}
-(\la+\frac{\cos d}{\la})&-q_x(0,y)+\frac{i\sin d}{\la}
\\
-q_x(0,y)-\frac{i\sin d}{\la}&\la+\frac{\cos d}{\la}
\end{array}\right).
\label{Vdef}\ee
We seek a non singular matrix $R(\la)$, independent of $y$, such that
\be
V(y,\frac 1 \la)=R(\la)V(y,\la)R(\la)^{-1}.
\label{conj}\ee
It can be verified that such a matrix is given by
\be
R(\la)=\left(\begin{array}{cc}
1&-F(\la)
\\
-F(\la)&1
\end{array}\right),
\label{Rdef}\ee
where $F$ is defined by (\ref{Fdef}).

Replacing in equation (\ref{phi2V}) $\la$ by $\frac 1 \la$,  and using (\ref{conj}),  we find the following equation:
$$
\left(R(\la)^{-1}\phi_2(y,\frac 1 \la)\right)_y =V(y,\la)\left(R(\la)^{-1}\phi_2(y,\frac 1 \la)\right),
$$
hence
$$
R(\la)^{-1}\phi_2(y,\frac 1 \la)=\phi_2(y,\la)C(\la),
$$
where $C$ is a $y$-independent matrix. Using the second of equations (\ref{ic}), it follows that $C=R^{-1}$, and therefore
$$
\phi_2(y,\frac 1 \la)=R(\la)\phi_2(y,\la)R(\la)^{-1}.
$$
This equation and equation (\ref{Fi2}) imply
\be
\Phi_2(y,\frac 1 \la)=R(\la)\Phi_2(y,\la)\left(\re^{\omega(\la)\frac {\widehat{\sigma_3}}{2}(y-L)}R(\la)^{-1}\right).
\label{mainconj}\ee
The first column vector of this equation implies (\ref{symm2}).

{\bf QED}

\begin{rem}
Recalling that $a_2(\la)=A_2(0,\la)$, and $b_2(\la)=B_2(0,\la)$, equations (\ref{symm2}) immediately imply the following
important relations:
\bea
&&a_2(\frac 1 \la)=\frac 1 {1-F(\la)^2}[a_2(\la)-F(\la)b_2(\la)+F(\la)\re^{-\omega(\la)L}b_2(-\la)-
F(\la)^2\re^{-\omega(\la)L}a_2(-\la)],
\nonumber \\
&&b_2(\frac 1 \la=\frac 1 {1-F(\la)^2}[b_2(\la)-F(\la)a_2(\la)+F(\la)\re^{-\omega(\la)L}a_2(-\la)-
F(\la)^2\re^{-\omega(\la)L}b_2(-\la)],
\nonumber \\
&&Im(\la)\geq 0.\quad \label{symm3}\eea
\end{rem}

In summary, the basic equations characterizing the spectral functions are:
\begin{itemize}
\item[(a)] the symmetry relations (\ref{symm1}) and (\ref{symm3});
\item[(b)] the global relations (\ref{grab1a}) and (\ref{grab2a});
\item[(c)] the conditions of unit determinant.
\end{itemize}
It turns out that, using these equations, it is possible to provide an explicit characterization of all the spectral functions in terms of the given constant $d$.

\begin{prop}
Assume that the functions $\{a_j(\la),b_j(\la)\}$, $j=1,2,3$ satisfy the symmetry relations (\ref{symm1}) and (\ref{symm3}), the global relations (\ref{grab1a}) and (\ref{grab2a}) and the "unit determinant" conditions
\be
a_j(\la)a_j(-\la)-b_j(\la)b_j(-\la)=1,\qquad j=1,2,3.
\label{dc}\ee
Then the following relations are valid:
\bea
&a_1(\la)b_1(-\la)-a_1(-\la)b_1(\la)=G(\la),&\qquad \la\in\R,
\label{rel1}\\
&a_3(\la)b_3(-\la)-a_3(-\la)b_3(\la)=-G(\la),&\qquad \la\in\R,
\label{rel2}\\
&a_2(\la)=a_1(-\la)a_3(\la)-\re^{-\omega(\la)L}b_1(\la)b_3(-\la),&\qquad \la\in\C,
\label{rel3}\\
&b_2(\la)=a_1(-\la)b_3(\la)-\re^{-\omega(\la)L}b_1(\la)a_3(-\la),&\qquad \la\in\C.
\label{rel4}
\eea
where
\be
G(\la)=\frac{i(1-\la^2)}{1+\la^2}\frac {\re^{\omega(\la)L}+\re^{-\omega(\la)L}-2}{\re^{\omega(\la)L}-\re^{-\omega(\la)L}}
\tan\frac d 2.
\label{Gdef}\ee
\end{prop}

\begin{rem} The two relations (\ref{rel3}) and (\ref{rel4}) are a direct consequence of the global relation and of the conditions of unit determinant. On the other hand, equations (\ref{rel1}) and (\ref{rel2}) depend  on the particular symmetry properties.
\end{rem}

{\bf Proof:} for simplicity,  we will use the notations
\be
f=f(\la), \qquad \hat f=f(-\la).\label{not1}\ee
Replacing $\la$ with $-\la$ in (\ref{grab1a}) and solving the resulting equation and equation (\ref{grab2a}) for $a_2(\la)$ and $b_2(\la)$ we find equations (\ref{rel3}) and (\ref{rel4}).

It can be verified directly that if $\{a_2,b_2\}$ are defined by equations (\ref{rel3}) and (\ref{rel4}) and $\{a_j,b_j\}$, $j=1,3$ satisfy the determinant condition (\ref{dc}), then $\{a_2,b_2\}$ also satisfies the determinant condition.

Replacing in the global relations(\ref{grab1a}) and (\ref{grab2a})  $\la$ by $-\frac 1 \la$, and using in the resulting equations the symmetry relations (\ref{symm1}), we find
\bea
&&a_1(\la)=a_2(\frac 1 \la)a_3(\la)-b_2(\frac 1 \la)b_3(\la),
\\
&&b_1(\la)\re^{\omega(\la)L}=a_2(-\frac 1 \la)b_3(\la)-b_2(-\frac 1 \la)a_3(\la).
\eea
Replacing in these equations $a_2(\pm\frac 1 \la)$ and $b_2(\pm\frac 1 \la)$ by the right hand side of the symmetry relation (\ref{symm3}), as well as by the right hand side of the equation obtained from equations (\ref{symm3}) under the transformation $\la\to-\la$,  after extensive simplifications we find the following equations, valid for $\la\in\R$:
\bea
\frac{(1-F^2)a_1}{b_1-Fa_1}=(a_3^2-b_3^2)\frac{\hat a_1-F\hat b_1}{b_1-Fa_1}+\Theta,
\\
\frac{(1-F^2)b_1}{a_1-Fb_1}=(a_3^2-b_3^2)\frac{\hat b_1-F\hat a_1}{a_1-Fb_1}+\Theta,
\eea
where
$$
\Theta=\re^{-\omega(\la)L}(b_3\hat a_3-\hat b_3a_3)+\re^{-\omega(\la)L}F,\qquad \la\in\R.
$$
Hence
\be
\frac{(1-F^2)a_1+\Delta(\hat a_1-F\hat b_1)}{b_1-Fa_1}=\frac{(1-F^2)b_1+\Delta(\hat b_1-F\hat a_1)}{a_1-Fb_1}=\Theta,\qquad \Delta=b_3^2-a_3^2.
\label{1stmess}\ee
Multiplying the numerator and denominator of the first fraction in (\ref{1stmess}) by $F$ and adding the resulting expression to the second fraction, as well as multiplying the numerator and denominator of the second fraction in (\ref{1stmess}) by $F$ and adding the resulting expression to the first fraction, we find
\be
\frac{Fa_1+b_1+\Delta\hat b_1}{a_1}=\frac {a_1+Fb_1+\Delta\hat a_1}{b_1}=\Theta.
\label{2ndmess}\ee
Using equation (\ref{dc}) with $j=1$, the left hand side of (\ref{2ndmess}) implies
$$
b_1^2-a_1^2=b_3^2-a_3^2.
$$
Then, the second equation of (\ref{1stmess}) implies
\be
\frac {a_1}{b_1}+\frac{\hat a_1(b_1^2-a_1^2)}{b_1}=\re^{-\omega(\la)L}(b_3\hat a_3-\hat b_3a_3)+\re^{-\omega(\la)L}F-F.
\label{3rdmess}\ee
The left hand side of (\ref{3rdmess}), using (\ref{dc}) with $j=1$, simplifies as follows:
$$
\frac {a_1}{b_1}+\hat a_1b_1-\frac {a_1}{b_1}(1+b_1\hat b_1)=\hat a_1b_1-a_1\hat b_1.
$$
Thus, equation (\ref{3rdmess}) becomes
\be
b_1\hat a_1-\hat b_1a_1=\re^{-\omega(\la)L}(b_3\hat a_3-\hat b_3a_3)+\re^{-\omega(\la)L}F-F.
\label{4thmess}\ee
Replacing in this equation $\la$ by $-\la$ yields
\be
-(b_1\hat a_1-\hat b_1a_1)=-\re^{\omega(\la)L}(b_3\hat a_3-\hat b_3a_3)+\re^{\omega(\la)L}F-F.
\label{5thmess}\ee
Equations (\ref{4thmess}) and (\ref{5thmess}), taking into account the definition (\ref{Fdef}) of $F$,  yield equations (\ref{rel1}) and (\ref{rel2}).

{\bf QED}
\begin{rem}
{\em
The determinant condition (\ref{dc}) with $j=1$ and equation (\ref{rel1}) imply
\be
\left[(a_1(\la)^2-b_1(\la)^2)\right]\left[(a_1(-\la)^2-b_1(-\la)^2)\right]=1-G(\la)^2,\quad \la\in\R.
\label{fact0}\ee
Indeed, equation (\ref{dc}) with $j=1$ implies the identity
\be
\left[(a_1^2-b_1^2)\right]\left[(\hat a_1^2-\hat b_1^2)\right]=1-(a_1\hat b_1-\hat a_1b_1)^2,\quad \la\in\R.
\label{fact01}\ee
Then equation (\ref{rel1}) implies (\ref{fact0}).}
\end{rem}

Equation (\ref{fact0}) defines the jump relation of a scalar RH problem for the sectional analytic functions
defined by
$$
\{(a_1(\la)^2-b_1(\la)^2,\;\la\in\C^+\;\quad a_1(-\la)^2-b_1(-\la)^2,\;\la\in\C^-\}.
$$
Taking into consideration that $a_1(\la)\neq b_1(\la)$ for $\la\in\C^+$ (otherwise equation (\ref{dc}) with $j=1$ is violated) it follows that the above Riemann-Hilbert problem has a unique solution
\be
a_1(\la)^2-b_1(\la)^2=h(\la),\qquad \la\in\C^+,
\label{sol1}\ee
where
\be
h(\la)=\re^{H(\la)}, \quad H(\la)=\frac 1 {2\pi i}\int_{\R}\ln[1-G^2(\la')]\frac{d\la'}{\la'-\la},\qquad \la\in\C\setminus\R.
\label{rhH}\ee
Using the fact that $G(\la)$ is an odd function, it follows that $H(\la)$ is also an odd function, hence
$h(-\la)=\re^{-H(\la)}$. This implies that the function $h(\la)$ defined by (\ref{rhH}) satisfies the jump condition (\ref{fact0}).

\begin{rem}
{\em
Equation (\ref{dc}) with $j=1$, and equations (\ref{rel1}) and (\ref{sol1}) imply
\be
a_1(-\la)=\frac{1}{h(\la)}\left(a_1(\la)+G(\la)b_1(\la)\right),\quad
b_1(-\la)=\frac{1}{h(\la)}\left(b_1(\la)+G(\la)a_1(\la)\right),\quad \la\in\R.
\label{sol2}\ee
Indeed, equation (\ref{rel1}) yields
$$
\hat b_1=\frac 1 {a_1}(G+\hat a_1 b_1).
$$
Replacing $\hat b_1$ in equation (\ref{dc}) with $j=1$  by the above expression, and making use of (\ref{sol1}), we find the first of equations (\ref{sol2}). The second of equations (\ref{sol2}) can be obtained in a similar way by eliminating $\hat a_1$ instead of $\hat b_1$. }
\end{rem}

\begin{rem}
{\em
The equations satisfied by $a_3$ and $b_3$ can be obtained from equations (\ref{sol2}) by replacing $G(\la)$ by $G(-\la)$. Hence
\be
a_3(-\la)=\frac{1}{h(\la)}\left(a_3(\la)-G(\la)b_3(\la)\right),\quad
b_3(-\la)=\frac{1}{h(\la)}\left(b_3(\la)-G(\la)a_3(\la)\right),\quad \la\in\R,
\label{sol2a3}\ee
where $G$ is given by (\ref{Gdef}) and $h(\la)$ is given by (\ref{rhH}).
}
\end{rem}

\begin{rem}
{\em
The function  $G$ is an entire function, thus each of equations (\ref{sol2}) defines the jump condition of a scalar RH problem.
However, it will be shown in section 5 that equations (\ref{sol2}) and (\ref{sol2a3}) are sufficient to determine the jump matrix (\ref{alljumps}). }
\end{rem}

\begin{rem}
{\em
Equations  (\ref{rel1})-(\ref{rel4}) imply the following identity:
\be
\re^{\omega(\la)L}[a_1(\la)^2-b_1(\la)^2]+\re^{-\omega(\la)L}[a_1(-\la)^2-b_1(-\la)^2]=
(\re^{\omega(\la)L}+\re^{-\omega(\la)L})(1-F^2)+2F^2,\quad \la\in\R.
\label{finid}\ee

Indeed, equations (\ref{rel3})-(\ref{rel4}) imply
\be
\re^{\omega(\la)L}(a_2^2-b_2^2)=\re^{\omega(\la)L}\hat a_1^2(a_3^2-b_3^2)-
\re^{-\omega(\la)L}b_1^2(\hat a_3^2-\hat b_3^2)-2\hat a_1b_1(a_3\hat b_3-\hat a_3b_3)
\label{6thmess}
\ee
Replacing in this equation $\la$ by $-\la$, adding the resulting equation to equation (\ref{6thmess}) and using equation (\ref{4thmess}) we find
\be
\re^{\omega(\la)L}(a_2^2-b_2^2)+\re^{-\omega(\la)L}(\hat a_2^2-\hat b_2^2)=
(\re^{\omega(\la)L}+\re^{-\omega(\la)L})(a_1^2-b_1^2)(\hat a_1^2-\hat b_1^2)+
2(a_1\hat b_1-\hat a_1b_1)(a_3\hat b_3-\hat a_3b_3).
\label{7thmess})\ee
Using (\ref{fact0}),
the right hand side of (\ref{7thmess}) equals the following expression:
$$
(\re^{\omega(\la)L}+\re^{-\omega(\la)L})-(a_1\hat b_1-\hat a_1b_1)\left[(\re^{\omega(\la)L}+\re^{-\omega(\la)L})(a_1\hat b_1-\hat a_1b_1)-2(a_3\hat b_3-\hat a_3b_3)\right].
$$
Using equations (\ref{rel1}) and (\ref{rel2}) the last expression becomes the right hand side of (\ref{finid}).
}
\end{rem}

\section{Spectral theory in the linearisable case}
\setcounter{equation}{0}
In the case of  the linearisable boundary conditions (\ref{lbc}), it is possible to express $q(x,y)$ in terms of the solution of a RH problem whose jump matrices are computed explicitly in terms of the given constant $d$. Indeed, recall that the jump matrices of the basic RH problem of section \ref{basicrh} are defined as follows:
 $$
 J^{\pi/2}=\left(\begin{array}{cc}
1&
I(\la)\re^{-\theta(x,y,\lambda)}
\\
\\
0&
1
 \end{array}\right),\quad
 J^{3\pi/2}=\left(\begin{array}{cc}
1& 0
\\
\\
I(-\la)\re^{\theta(x,y,\lambda)}&1
 \end{array}\right),\qquad I(\la)=\frac{b_2(-\lambda)}{a_1(\la)a_3(\lambda)},
$$
$$
J^{0}=\left(\begin{array}{cc}
R(\la)& \frac {b_3(-\lambda)}{a_3(\lambda)}\re^{-\theta(x,y,\lambda)}
\\
\\
-\frac{\re^{-\omega(\lambda)L}b_1(\lambda)}{a_1(-\lambda)}\re^{\theta(x,y,\lambda)}& 1
 \end{array}\right),
 \qquad R(\la)=\frac {a_2(\lambda)}{a_1(-\lambda)a_3(\lambda)},$$
and
\be
J^{\pi}=
J^{3\pi/2}(J^{0})^{-1}J^{\pi/2},
\label{alljumps2}\ee
Equation (\ref{rel3}) and (\ref{rel4}) imply that
\be
R(\la)=1-\re^{-\omega(\la)L}\frac{\hat b_3}{a_3}\frac {b_1}{\hat a_1},
\quad
I(\la)=\frac {\hat b_3}{a_3}-\re^{\omega(\la)L}\frac{\hat b_1}{a_1}.
\label{IR}\ee

Thus in the linearisable case, the jump matrices involve only the rations $\frac {\hat b_3}{a_3}$ and $\frac{\hat b_1}{a_1}$, evaluated at $\la$ and at $-\la$. Equations (\ref{rhH}) and (\ref{sol1})  imply that these rations are given by
\be
\frac {\hat b_3}{a_3}=-\frac{G}{h}+\frac{b_3}{a_3h},\qquad
\frac {\hat b_1}{a_1}=\frac{G}{h}+\frac{b_1}{a_1h}.
\label{hatnohat}\ee
Hence the jump matrices depend on the known function $\frac Gh$ as well as on the unknown functions
$\frac{b_1}{a_1h}$ and $\frac{b_3}{a_3h}$. Using the fact that these unknown functions are bounded and analytic in $\C^+$, it is possible to formulate a RH problem, equivalent to the basic  one defined by (\ref{alljumps2}), in terms of the known function $\frac G h$ only. This new RH problem is therefore defined by the following jump matrices:
$$
\tilde  J^{\pi/2}=\left(\begin{array}{cc}
1&
\tilde I(\la)\re^{-\theta(x,y,\lambda)}
\\
\\
0&
1
 \end{array}\right),\quad
\tilde  J^{3\pi/2}=\left(\begin{array}{cc}
1& 0
\\
\\
\tilde I(-\la)\re^{\theta(x,y,\lambda)}&1
 \end{array}\right),\qquad \tilde I(\la)=-\frac{G}{h}(1+\re^{-\omega(\la)L}),
$$
$$
\tilde J^{0}=\left(\begin{array}{cc}
\tilde R(\la)&- \frac {G}{h}\re^{-\theta(x,y,\lambda)}
\\
\\
\frac{\re^{-\omega(\lambda)L}G}{\hat h}\re^{\theta(x,y,\lambda)}& 1
 \end{array}\right),
 \qquad \tilde R(\la)=1-\frac {G^2}{h\hat h}\re^{-\omega(\la)L},$$
and
\be
\tilde J^{\pi}=
\tilde J^{3\pi/2}(\tilde J^{0})^{-1}\tilde J^{\pi/2},
\label{alljumps2alt}\ee

\begin{thm}
Let $q(x,y)$ satisfy equation (\ref{esg}) and the boundary conditions (\ref{lbc}).


Then $q(x,y)$ is given by equations (\ref{qexp1})-(\ref{qexp2}) with $\Psi$ replaced by $\tilde\Psi$, where $\tilde\Psi$ is the solution of the Riemann-Hilbert problem (\ref{rh}) with the jump matrix $J$ replaced by the matrix $\tilde J$  defined as follows:
\be
\tilde J(x,y,\lambda)=
\tilde J^{\alpha}(x,y,\lambda),\;\;{\rm if}\;\;arg(\lambda)=\alpha,\quad
\alpha=0,\,\frac{\pi}{2},\, \pi,\,\frac{3\pi}{2},
\label{newjump} \ee
where
$$
\tilde J^{0}=\left(\begin{array}{cc}
1-\frac{G^2(\la)}{h(\la) h(-\la)}\re^{-\omega(\lambda)L}&-\frac{G(\la)}{h(\la)}\re^{-\theta(x,y,\lambda)}
\\
\\
\re^{-\omega(\lambda)L}\frac{G(\la)}{h(-\la)}\re^{\theta(x,y,\lambda)}& 1
 \end{array}\right),
$$
 $$
\tilde J^{\pi/2}=\left(\begin{array}{cc}
1&
-\frac{G(\la)}{h(\la)}\left(1+\re^{\omega(\lambda)L}\right)\re^{-\theta(x,y,\lambda)}
\\
\\
0&1
 \end{array}\right),\qquad
\tilde  J^{3\pi/2}=\left(\begin{array}{cc}
1& 0
\\
\\
\frac{G(\la)}{h(-\la)}\left(1+\re^{-\omega(\lambda)L}\right)\re^{\theta(x,y,\lambda)}&1
 \end{array}\right)
$$
and
\be
\tilde J^{\pi}=
\tilde J^{3\pi/2}(\tilde J^{0})^{-1}\tilde J^{\pi/2},
\label{alljumpslin}\ee
where $
\theta(x,y,\lambda)$ is given by
(\ref{Theta}), while
$G(\la)$
and $h(\la)$ are defined in terms of the given constant $d$ by equations (\ref{Gdef}) and (\ref{rhH}).

This \RH problem is regular and, if $d\in\R$,  it has a unique solution.
\end{thm}

{\bf Proof}:
The solution $\tilde\Psi$ of the new  RH problem satisfies the jump relation
 \be
\tilde\Psi_{-}(x,y,\lambda)=\tilde\Psi_{+}(x,y,\lambda)\tilde J(x,y,\lambda), \lambda\in\R\cup i\R,
\label{rhtilde}\ee
where  the jump matrix $\tilde J$ is given by equations (\ref{newjump}) and (\ref{alljumpslin}) (the latter are identical to equations (\ref{alljumps2alt})). Let $\Psi_j$ and $\tilde \Psi_j$, $j=1,...,4$, denote $\Psi$ and $\tilde\Psi$  in the $j$-th quadrant of the complex $\lambda$ plane, respectively. We seek matrices $A_j$, $j=1,...,4$ which are analytic and bounded in the $j$-th quadrant of the complex $\lambda$ plane, respectively, and such that
\be
 \tilde \Psi_j(x,y,\la)=\Psi_j(x,y,\la)A_j(x,y,\la),\quad (j-1)\frac{\pi}2< arg(\la)<j\frac{\pi}{2},\qquad j=1,...,4.
 \label{Aidef}
\ee
Let
  \bea
&& A_1=\left(\begin{array}{cc}
1&
\alpha_1(\la)\re^{-\theta(x,y,\lambda)}
\\
\\
0&1
 \end{array}\right),\qquad 0<arg(\la)<\frac\pi 2
 \label{A1}\\
&& A_2=\left(\begin{array}{cc}
1&
\alpha_2(\la)\re^{-\theta(x,y,\lambda)}
\\
\\
0&1
 \end{array}\right),\qquad\frac\pi 2<arg(\la)<\pi
  \label{A2}\\
 && A_3=\left(\begin{array}{cc}
1&0
\\
\\
\alpha_3(\la)\re^{\theta(x,y,\lambda)}&1
 \end{array}\right),\qquad\pi <arg(\la)<\frac{3\pi}{2}
 \label{A3} \\
&& A_4=\left(\begin{array}{cc}
1&0
\\
\\
\alpha_4(\la)\re^{\theta(x,y,\lambda)}&1
 \end{array}\right),\qquad\frac{3\pi} 2<arg(\la)<2\pi.
\label{A4}
 \eea
 Equations (\ref{rh}) and (\ref{rhtilde}) imply the following relations:
\be
  J^{\pi/2}A_2=A_1\tilde J^{\pi/2},
   \qquad
   J^{3\pi/2}A_4=A_3\tilde J^{3\pi/2},
   \qquad
  J^{0}A_4=A_1\tilde J^{0}.
   \label{psifinrel} \ee
   The first two equations (\ref{psifinrel}) imply
   \be
   \alpha_2(\la)+I(\la)=\alpha_1(\la)+\tilde I(\la),\qquad
   \alpha_4(\la)+I(-\la)=\alpha_3(\la)+\tilde I(-\la).
   \label{Ialpha}\ee
  Using the definitions of $I(\la)$ and $\tilde I(\la)$ (see equations (\ref{IR}) and (\ref{alljumps2alt}))
   as well as (\ref{hatnohat}), equations (\ref{Ialpha}) become
   $$
   \alpha_2+\frac{b_3}{a_3h}-\re^{\omega(\la)L}\frac{b_1}{a_1h}=\alpha_1, \qquad
   \alpha_4+\frac{\hat b_3}{\hat a_3\hat h}-\re^{-\omega(\la)L}\frac{\hat b_1}{\hat a_1\hat h}=\alpha_3.
   $$
   The simplest solution of these equations that satisfy the requirement that the functions $\alpha_j(\la)$ are bounded and analytic in the $j$-th quadrant of the complex $\la$-plane, $j=1,...,4$, are the following:
\bea
\alpha_1=
\frac{b_3}{a_3h},&& \alpha_2=
\frac{b_1}{a_1h}\re^{\omega(\la)L},\nonumber \\ \alpha_3=\frac{\hat b_3}{\hat a_3\hat h},
&& \alpha_4=\frac{\hat b_1}{\hat a_1\hat h}\re^{-\omega(\la)L}.\label{alphas}
\eea
The third of equations (\ref{psifinrel}) yields the following relations:
\bea
&&\frac{\hat b_3}{a_3}=\alpha_1-\frac{G}{h},\qquad
\frac{b_1}{\hat a_1}\re^{-\omega(\la)L}=\alpha_4-\frac G{\hat h}\re^{-\omega(\la)L} ,\nonumber
\\
&&R+\alpha_4\frac{\hat b_3}{a_3}=\tilde R+\alpha_1\re^{-\omega(\la)L}\frac G {\hat h}.
\label{lasteqs} \eea
Using the definitions of $R(\la)$, $\tilde R(\la)$ (equations (\ref{alljumps2}), (\ref{alljumps2alt})) the definitions of $\alpha_j$ , $j=1,..,4$ (equations (\ref{alphas})) and and equations (\ref{hatnohat}) it can be verified that equations (\ref{lasteqs}) are satisfied identically.

\smallskip

 The eigenfunctions $\tilde\Psi_j$, $j=1,...,4$ are given explicitly in terms of the eigenfunctions $\Psi_j$ by equations (\ref{Aidef}). This yields the following expressions:
 \bea
&& \tilde\Psi_1=(\Psi_1^{(12)}, \frac{b_3}{a_3h}\Psi_1^{(12)}\re^{-\theta(x,y,\la)}+\frac1 {a_3}\Psi_3^{(1)}),\quad
  \tilde\Psi_2=(\Psi_1^{(12)}, \frac{b_1}{a_1h}\re^{\omega(\la)L}\Psi_1^{(12)}\re^{-\theta(x,y,\la)}+\frac1 {a_1}\Psi_2^{(2)}),
  \nonumber \\
  \label{tildeef}\\
  &&\tilde\Psi_3=(\frac 1 {\hat a _3}\Psi_3^{(3)}+\frac{\hat b_3}{\hat a_3\hat h}\Psi_1^{(34)}\re^{\theta(x,y,\la)},\Psi_1^{(34)}),\quad
  \tilde\Psi_4=(\frac 1 {\hat a _1}\Psi_2^{(4)}+\frac{\hat b_1}{\hat a_1\hat h}\re^{-\omega(\la)L}\Psi_1^{(34)}\re^{\theta(x,y,\la)},\Psi_1^{(34)}).\nonumber
  \eea
The above equations can be simplified as follows:
 \bea
&& \tilde\Psi_1=(\Psi_1^{(12)}, \frac{b_3}{h}\Psi_3^{(3)}\re^{-\theta(x,y,\la)}+\frac{a_3}{h}\Psi_3^{(1)}),\quad
  \tilde\Psi_2=(\Psi_1^{(12)}, \frac{b_1\re^{\omega(\la)L}}{h}\Psi_2^{(4)}\re^{-\theta(x,y,\la)}+\frac{a_1}{h}\Psi_2^{(2)}),
  \nonumber \\
  \label{tildeef2}\\
  &&\tilde\Psi_3=(\frac{\hat a_3}{\hat h}\Psi_3^{(3)}+\frac{\hat b_3}{\hat h}\Psi_3^{(1)}\re^{\theta(x,y,\la)},\Psi_1^{(34)}),\quad
  \tilde\Psi_4=(\frac{\hat a_1}{\hat h}\Psi_2^{(4)}+\frac{\hat b_1\re^{-\omega(\la)L}}{\hat h}\Psi_2^{(2)}\re^{\theta(x,y,\la)},\Psi_1^{(34)}).\nonumber
  \eea

Indeed, using equation (\ref{jump3}) we find
$$
 \frac{b_3}{a_3h}\Psi_1^{(12)}\re^{-\theta(x,y,\la)}+\frac1 {a_3}\Psi_3^{(1)}=
  \frac{b_3}{a_3h}(a_3\Psi_3^{(3)}+b_3\Psi_3^{(1)}\re^{\theta(x,y,\la)})\re^{-\theta(x,y,\la)}+\frac1 {a_3}\Psi_3^{(1)}
  $$
  Using $a_3^2-b_3^2=h$ the above expression is equal to the regular function
  $$
   \frac{b_3}{h}\Psi_3^{(3)}\re^{-\theta(x,y,\la)}+\frac{a_3}{h}\Psi_3^{(1)}.
   $$
  A similar computation yields the result for the other eigenfunctions.

Thus the above \RH problem (\ref{rhtilde}) is regular.

\smallskip

We now prove that the \RH problem defined by (\ref{rhtilde}) is uniquely solvable. It can be verified that when $d\in\R$, then $h(\la)=\overline{h(\bar \la)}$. In this case, the jump matrices $\tilde J^{(\alpha)}$ satisfy the following conditions: the matrices are Schwarz invariant on the imaginary axis and have zero real part on the real axis of the complex $\la$ plane. Under these assumptions,  it follows from general results ( see e.g. \cite{deift, zhoufok, zhou}) that the so-called ``vanishing lemma" holds. This guarantees the existence of a  unique solution.

{\bf QED}

\section{Conclusions}
We have studied boundary value problems for the elliptic sine-Gordon posed on a semistrip.
In particular we have shown that if the prescribed boundary conditions are  zero along the unbounded sides of the semistrip and constant on the bounded side, then it is possible to obtain the solution in terms of a Riemann-Hilbert problem which is uniquely defined in terms of   the width $L$ of the semistrip and the constant value $d$ of the solution along the $x=0$ boundary.  Indeed,  the ``jump matrices" of this \RH problem  are defined in terms of the two functions  $G(\la)$ and $h(\la)$ which are explicitly defined, by equations (\ref{Gdef}) and (\ref{rhH}) respectively, in terms of $L$ and $d$.
Furthermore, this \RH problem has a unique solution, see theorem 5.1.

\section*{Acknowledgements}
This research was partially  supported by EPSRC grant EP/E022960/1. ASF would like to express his gratitude to the Guggenheim Foundation, USA.

\end{document}